# Upgrade of the Minos+ Experiment Data Acquisition for the High Energy NuMI Beam Run

William Badgett, Steve R. Hahn, Donatella Torretta, Jerry Meier, Jeffrey Gunderson, Denise Osterholm, David Saranen

*Abstract*–The Minos+ experiment is an extension of the Minos experiment at a higher energy and more intense neutrino beam, with the data collection having begun in the fall of 2013. The neutrino beam is provided by the Neutrinos from the Main Injector (NuMI) beam-line at Fermi National Accelerator Laboratory (Fermilab). The detector apparatus consists of two main detectors, one underground at Fermilab and the other in Soudan, Minnesota with the purpose of studying neutrino oscillations at a base line of 735 km. The original data acquisition system has been running for several years collecting data from NuMI, but with the extended run from 2013, parts of the system needed to be replaced due to obsolescence, reliability problems, and data throughput limitations. Specifically, we have replaced the front-end readout controllers, event builder, and data acquisition computing and trigger processing farms with modern, modular and reliable devices with few single points of failure. The new system is based on gigabit Ethernet TCP/IP communication to implement the event building and concatenation of data from many front-end VME readout crates. The simplicity and partitionability of the new system greatly eases the debugging and diagnosing process. The new system improves throughput by about a factor of three compared to the old system, up to 800 megabits per second, and has proven robust and reliable in the current run.

## I. INTRODUCTION

THE Minos experiment [1] has been operating for several years in the NuMI (Neutrinos from the Main Injector) muon neutrino beam produced at Fermi National Accelerator Laboratory (Fermilab). The primary physics goal of Minos is to study muon neutrino oscillations between the near and far detectors, where the muon neutrino oscillates into another type of neutrino between Fermilab and the far detector at Soudan, Minnesota, 735 km away.

The Minos detectors consist of steel absorber plates alternating with scintillator readout layers arranged in orthogonal directions. The detectors provide both a calorimeter energy measurement and tracking via the scintillator segmentation. Solenoids provide a magnetic field for the measurement of muon momentum in the case of neutrino charge current interactions. See Fig. 3 for a photograph of the Minos Far Detector, shown on the left side of the image.

Minos was designed a number of years before data taking commenced; by dint of technical progress, some of the electronics technology involved has become obsolete. The Minos+ experiment was approved to extend the Minos data set in the Nova [2] NuMI beam era [3] with a higher energy and more intense neutrino beam. Both improved beam features result in higher signal occupancies in the Minos detector, and especially so at the near detector located at Fermilab. The near detector is situated at 371 meters from the decay pipe, while the far detector is 735 km away. At the peak beam intensities of the high-energy run, we expect twenty neutrino interactions at the near detector per spill of ten micro-seconds duration at the near detector, compared to a maximum of three in the run ending in 2011. Also for the high-energy run, the spill repetition period decreases from 1.9 seconds to 1.33 seconds, further taxing the existing system.

## II. THE MINOS DATA ACQUISITION SYSTEM

The Minos data acquisition front-end electronics differ in the near and far detector, but both systems operate in a very similar fashion, the general features we describe here. Both are described extensively in detail in reference [1]. The electronics run continuously at a sampling frequency of approximately 53 MHz on a clock synchronized to several GPS satellites, storing for each sample the collected charge on one capacitor in an array of capacitors. On the receipt of a trigger, the selected capacitor's charge is digitized and sent to a digital buffer card, along with the determined time stamp and trigger type. The digital buffer cards are VME based and implement a double buffering system. Live signals will be recorded in one of the buffers, while previously recorded signals in the other buffer are read via the VME bus by a VME crate controller. See Fig. 1 for a photograph of the near detector front-end electronics.

One major difference between near and far detectors is that the near detector has direct access to the NuMI beam accelerator timing signals. During a window around the ten micro-second beam spill, the near detector electronics digitizes all time samples. Between spills, multiple channels over threshold are required to trigger time samples, allowing an efficient cosmic ray trigger that does not overwhelm the data throughput. For the far detector, the front-end continuously runs with the channels over threshold trigger requirement. The beam spill time stamp is sent via a network packet from near to far; due to network latency, this information is only used in the trigger processor farm downstream of front-end data collection.

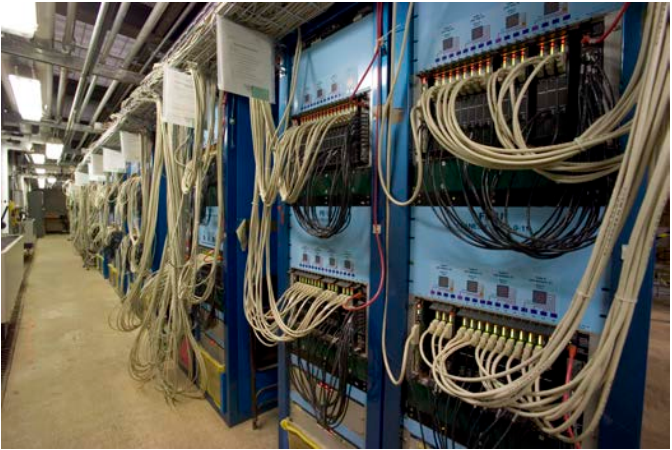

Fig. 1: Minos Near Detector Data Acquisition Electronics

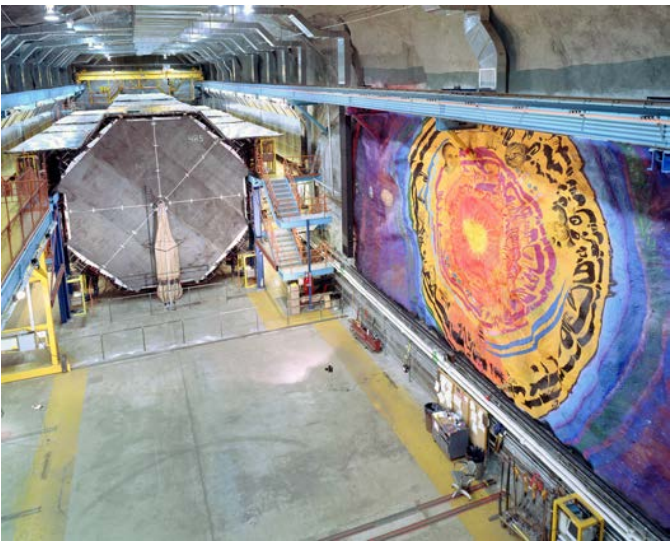

Fig. 2 The Minos Far Detector

### III. THE ORIGINAL MINOS EVENT BUILDER

The Minos event builder involves reading out the digital buffer boards, collecting the data across front-end crates and serving the result to the trigger processing farm. A diagram of the original Minos event builder is shown in Fig. 3. The original system used the PVIC [4] data link technology, implementing a complex topology. The CES/RIO [4] VME front-end crate processors were daisy-chained in pairs or triplets. Fig. 3 shows the near detector topology; the far detector is a bit more complex with sixteen total front-end VME crates.

The BRP (branch processor) computers collect data from two (near) or three (far) front-end crates. The data are then collected across the several BRP computers and sent onto a PVIC bus shared by the TP (trigger processor farm) and BRP computers and assembled into a single event data frame. The TP farm analyzes and selects a subset of data for final output on disk on separate data logging computers.

This PVIC based event builder was able to handle a maximum of approximately 200 Mb/sec (million bits per second) of throughput which would not be adequate for the higher data rates expected during the high-energy, high-intensity NuMI beam run starting in 2013. The PVIC design also presented several single points of failure, a difficulty of partitioning and a design complexity which led to a time consuming process of diagnosing problems with the system. By 2012, the PVIC interfaces and drivers were obsolete and difficult or impossible to replace or repair, and the high rate of failures necessitated a full replacement.

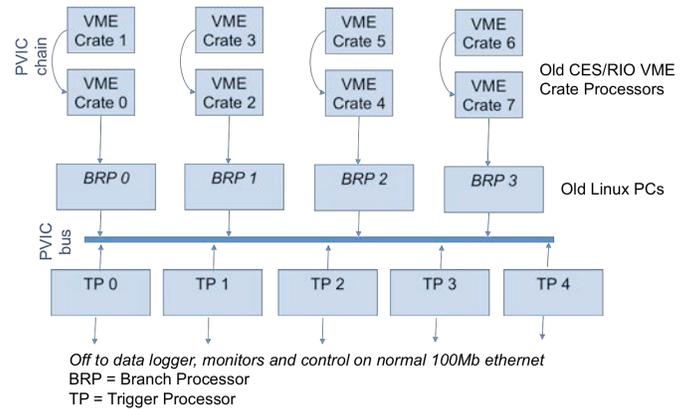

Fig. 3: Original Minos Event Builder

### IV. THE NEW MINOS EVENT BUILDER

The goals for the new event builder were:

- Increase the data though-put to at least 800 Mb/sec
- Implement a modular and easily partitioned architecture
- Minimize single points of failure
- Use commodity, standard commercial hardware where possible
- Consolidate the data acquisition functions to fewer computers
- Update all operating systems to their latest versions
- Remove all PVIC devices and their old servers
- Make sure computers are interchangeable and not dedicated to a single process
- Reduce, reüse, recycle

The diagram for the chosen design can be seen in Fig. 5. The PVIC data flow has been replaced by standard gigabit Ethernet, using a private, dedicated local area network (LAN) to transport the detector data, and only that data. Data are sent from the front-end VME crate processors over the network to a single BRP running on an eight-processor Linux server running Scientific Linux 6.3 [5]. Taking advantage of the increased number of processor cores on newer computers, we implement the BRP to TP communications via local message passing. The use of shared memory buffers allows multiple processes to access the data with minimal delay and no data transfer penalty.

Fig. 4 shows the installation at the near detector of one MVME5500 [6] VME crate processor in the readout digital buffer crates. Note the three links connecting to the front-panel, one for the data transfer network (green), one for the local control network (blue) and one for the serial boot console (white). As shown in Fig. 5, the data and control networks are distinct private LANs in order to maximize the data throughput. All control, status, configuration and interactive logins go through the lower speed control network connection (blue). The white serial boot console connections are provided for boot problem diagnosing and programming the network boot parameters. The MVME5500 brings greater processing power at 1 GHz clock speed (compare 400 MHz for CES/RIO) and one gigabyte of memory, allowing greater data caching buffer sizes.

The data acquisition tasks running under VxWorks [7] on the MVM5500 have been substantially rewritten to adapt to the new hardware, upgrading to the new, modern VxWorks version 6.9 and to implement the new gigabit Ethernet data transport. The resulting data throughput can now reach up to 800 Mb/sec. With the simplified architecture, each VME crate readout can easily be isolated for simple diagnosis of system problems. At each stage in Fig. 5 we have implemented multiple buffer caching schemes in order to handle the inherent burstiness of our periodic beam spill environment.

Every 50 milli-seconds the MVME5500 handles an interrupt from the central timing system, synchronized to the GPS. The interrupt initiates swapping the buffer on the digital data buffers cards, and reading out the ready data buffer. At the end of one-second duration (or twenty 50 milli-second buffer swaps) the MVME5500s send their data to the single BRP Branch Processor over the private dedicated network.

At the BRP and TP level, all processes run on the latest Scientific Linux [5] with modern shared memory handling, in contrast to the hardware specific PVIC hard-wired memory locations. The BRP collates all eight (near) or sixteen (far) time frame data packets from the front-end MVME550 processors. With the Linux eight-processor server and common shared memory, the data transport mechanism is greatly simplified and the data transport speed from BRP to TP is limited only by the local control message transport latency in the TCP/IP *localhost* protocol. The TP trigger processor processes and analyzes the detector data, filtering out unwanted background data hits, selecting only potential physics interaction data. The TP processes run as a farm of up to five instances, with the BRP assigning a TP process in a sequential basis. Current data rates tend to require on average two instances.

Downstream from the TP processes are data formatters, data loggers and online monitoring applications running on a separate server not shown directly on Fig. 5. All of these processes run on a similar, modern, eight-processor server identical to the BRP and TP server. All of these applications have been brought up to 2014 standards. The third remaining data acquisition server hosts the run control communication server process, and various calibration control and online status monitoring processes. The data acquisition servers have now been reduced from sixteen to three.

## V. Conclusion

We have implemented a new event builder data transport and concatenation system for the two Minos+ detectors in anticipation of the high-energy, high-intensity NuMI beam run. The new system is modular with very few single points of failure. The increased reliability of the new system allows the data acquisition to run without human intervention for weeks, and no hardware repairs or replacements have so far been necessary. The elegance of the new system even allowed an emergency switch over from the old to the new system at the far detector within thirty-six hours in February 2014.

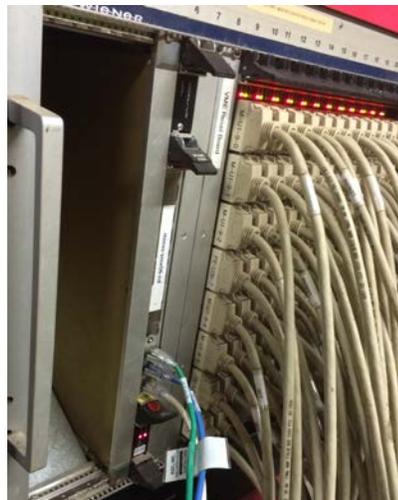

Fig. 4: Minos Near Detector Digital Buffer Crate with MVME5500

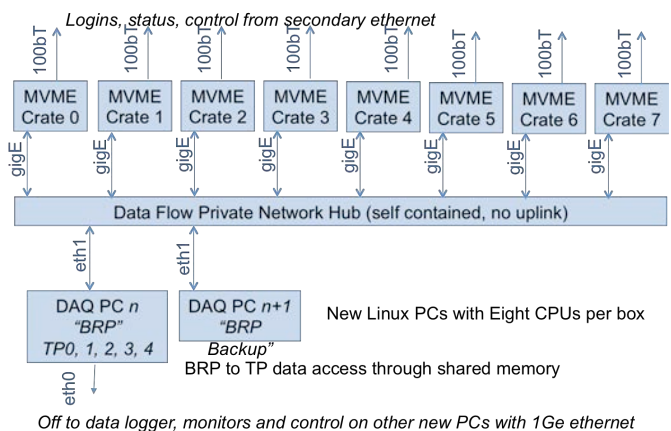

Fig. 5: New Minos Event Builder